\documentclass[twocolumn,aps,prl,floatfix,amsmath,amssymb,superscriptaddress,longbibliography]{revtex4-2}

\usepackage{wasysym}
\usepackage{graphicx}
\usepackage{dcolumn}
\usepackage{bm}
\usepackage{multirow}
\usepackage{amssymb}
\usepackage{graphicx}
\usepackage{braket}
\usepackage{xcolor}
\definecolor{TighnariBrown}{RGB}{141,87,41}
\definecolor{TighnariGreen}{RGB}{40,114,70}
\definecolor{TighnariYellow}{RGB}{247,191,99}
\definecolor{PRLBlue}{RGB}{46,48,146}
\usepackage[
    colorlinks,
    citecolor=PRLBlue,
    linkcolor=PRLBlue,
    urlcolor=PRLBlue,
]{hyperref}
\usepackage{cases}
\usepackage{pifont}
\newcommand{\cmark}{\ding{51}}
\newcommand{\xmark}{\ding{55}}

\makeatletter
\newcommand{\equalcontrib}{\frontmatter@footnote{These two authors contributed equally to this work.}}
\makeatother

\begin{document}
\title{Nonstabilizerness Mpemba Effects}

\author{Zhenyu Xiao\equalcontrib}
\email{zyxiao@princeton.edu}
\affiliation{Princeton Quantum Initiative, Princeton University, Princeton, New Jersey 08544, USA}

\author{Hao-Kai Zhang}
\email{hkzhang@iphy.ac.cn}
\affiliation{Institute of Physics, Chinese Academy of Sciences, Beijing 100190, China}

\author{Shuo Liu\equalcontrib}
\email{sl6097@princeton.edu}
\affiliation{Department of Physics, Princeton University, Princeton, New Jersey 08544, USA}

\date{\today}

\begin{abstract}
Quantum state preparation can be strikingly counterintuitive: the fastest route to a target state need not start from the apparently closest initial condition.
We uncover such a quantum Mpemba effect in the dynamical generation of quantum magic (nonstabilizerness), quantified by the stabilizer R\'enyi entropy, in $\mathrm{U(1)}$-symmetric random circuits initialized from tilted product states. 
States with lower initial magic can generate magic faster than states with higher initial magic.
The acceleration is not determined solely by the conserved-charge distribution. 
Two initial-state families with identical initial magic and identical charge distribution exhibit qualitatively different magic-growth dynamics, depending also on the spatial structure of the initial state within each charge sector.
Analogous magic Mpemba effects in $\mathrm{SU(2)}$-symmetric circuits and under nonintegrable Hamiltonian dynamics further show that the phenomenon is tied neither to Abelian symmetry nor to random-circuit dynamics, establishing quantum magic as a distinct arena for Mpemba physics.
\end{abstract}

\maketitle

\textbf{Introduction.---}
Quantum magic, also known as nonstabilizerness, is a central resource for quantum computation~\cite{BravyiKitaev2005,Veitch2014Resource, ContextualitySuppliesTheMagicForQuantumComputation, HowardCampbell2017}.
Although Clifford circuits acting on stabilizer states can generate extensive entanglement, they remain efficiently classically simulable~\cite{PhysRevA.70.052328,gottesman1998heisenbergrepresentationquantumcomputers}.
Escaping this simulable regime requires nonstabilizer resources: magic quantifies precisely this departure from stabilizer structure and supplements Clifford operations to enable universal and fault-tolerant quantum computation~\cite{shor1996fault, PhysRevA.57.127, aharonov1997fault,gottesman2010introduction}.
It is closely connected to quasiprobability negativity and contextuality~\cite{Veitch2012Negative,Veitch2014Resource,BermejoVega2017}, governs the overhead of classical simulation~\cite{QuantifyingQuantumSpeedupsImprovedClassicalSimulationFrom}, and is the resource refined by magic-state distillation~\cite{MagicStateDistillationWithLowOverhead,MagicStateDistillationWithLowSpaceOverhead,MagicStateDistillationNotAsCostlyAs}.
How magic is dynamically generated in many-body systems is therefore central to quantum state preparation~\cite{PhysRevResearch.3.043200,PhysRevLett.129.230504,10044235,rosenthal2023querydepthupperbounds,Yuan2023optimalcontrolled,Low2024tradingtgatesdirty,Zhang2024parallelquantum}, and recent progress in scalable magic monotones and many-body diagnostics~\cite{Leone2022SRE,StabilizerEntropiesAndNonstabilizernessMonotones,ScalableMeasuresOfMagicResourceForQuantum,QuantumMagicDynamicsInRandomCircuits,MagicSpreadingInRandomQuantumCircuits,CertifyingNonstabilizernessInQuantumProcessors} has made such a study feasible~\cite{tirrito2025anticoncentrationnonstabilizernessspreadingergodic,aditya2025growthspreadingquantumresources,Magni2025quantumcomplexity,turkeshiMagicSpreadingRandom2025a,xfp5-hhs4,195d-r5j3}, raising a natural question: how can nonstabilizer resources be generated most efficiently?

On a different front, the Mpemba effect~\cite{MpembaOsborne1969}---namely, the counterintuitive phenomenon that hot water may cool faster than cold water under otherwise identical conditions---has attracted sustained attention. 
In classical settings, it has been explored, for example, in water-freezing and cooling protocols~\cite{Kell1969Freezing,Auerbach1995Supercooling,Katz2009HotWater,Jeng2006Mpemba,Brownridge2011Search,VynnyckyKimura2015Convection,BurridgeLinden2016Questioning}, stochastic Markov dynamics and anomalous relaxation~\cite{LuRaz2017Markovian,KlichRazHirschbergVucelja2019}, granular media~\cite{Lasanta2017Granular}, spin glasses~\cite{BaityJesi2019SpinGlasses}, colloidal relaxation experiments~\cite{KumarBechhoefer2020Cooling}, and related classical settings~\cite{HJXMeanField,HJXMeanfield2,HJXFirstOrderPT,HJXOrderParameter,lin2026macroscopicmpembaeffectcumulativeheatenhanced}. Despite the extensive work, even its precise scope and robustness remain under active debate~\cite{Jeng2006Mpemba,Brownridge2011Search,VynnyckyKimura2015Convection,BurridgeLinden2016Questioning}. 
The Mpemba effect has more recently been generalized to quantum settings~\cite{TheQuantumMpembaEffects,QuantumMpembaEffectsFromSymmetryPerspectives, SpeedupsInNonequilibriumThermalRelaxation,TheQuantumMpembaEffectInClosedSystemsFromTheoryToExperiment,ResourceTheoreticalUnificationOfMpembaEffects}. 
In open quantum systems, the effect manifests as accelerated relaxation toward nonequilibrium steady states or thermal stationary states~\cite{LindbladDissipativeDynamicsInThePresenceOf,ExponentiallyAcceleratedApproachToStationarityInMarkovian,AcceleratingTheApproachOfDissipativeQuantumSpin,QuantumMpembaEffectInAQuantumDot,HyperaccelerationOfQuantumThermalizationDynamicsByBypassing,ThermodynamicsOfTheQuantumMpembaEffect,AharonyShapira2024,MultipleQuantumMpembaEffectExceptionalPointsAnd,MpembaEffectsInNonequilibriumOpenQuantumSystems,MpembaEffectsInOpenNonequilibriumQuantumSystems,PhotonicMpembaEffect,BosonicMpembaEffectWithNonClassicalStates,SpeedingUpQuantumHeatEnginesByThe,EntangledMultipletsAsymmetryAndQuantumMpembaEffect,DynamicalInvariantBasedShortcutToEquilibrationIn,EnhancedQuantumMpembaEffectWithSqueezedThermal,IntrinsicQuantumMpembaEffectInMarkovianSystems,QuantumMpembaEffectOfLocalizationInThe,TheMpembaEffectInQuantumOscillatingAnd,NonMarkovianQuantumMpembaEffect,QuantumMpembaEffectFromInitialSystemReservoir,QuantumMpembaEffectInParityTimeSymmetricSystems,NoiseInducedQuantumMpembaEffect,MpembaEffectAndSuperAcceleratedThermalizationInTheDampedQuantumHarmonicOscillator,QuantumMpembaEffectFromNonNormalDynamics,AGeneralStrategyForRealizingMpembaEffectsInOpenQuantumSystems,CanonicalQuantumMpembaEffectInADissipativeQubit,MpembaMeetsQuantumChaosAnomalousRelaxationAnd,EngineeringQuantumMpembaEffectByLiouvillianSkinEffect,QuantumMpembaEffectInDissipativeSpinChainsAtCriticality}.
In isolated quantum systems, by contrast, closely related anomalous relaxation phenomena arise as symmetry restoration, where more asymmetric initial states can restore the symmetry faster~\cite{EntanglementAsymmetryAsAProbeOfSymmetry,LiuZhangYinZhang2024,QuantumMpembaEffectsInManyBodyLocalization,LackOfSymmetryRestorationAfterAQuantum,EntanglementAsymmetryInTheOrderedPhaseOf,Rylands2024,Joshi2024,EntanglementAsymmetryAndQuantumMpembaEffectIn,MultipleCrossingsDuringDynamicalSymmetryRestorationAnd,EntanglementAsymmetryAndQuantumMpembaEffectInX,DynamicalSymmetryRestorationInTheHeisenbergSpin,NonEquilibriumEntanglementAsymmetryForDiscreteGroups,AUniversalFormulaForTheEntanglementAsymmetry,RenyiEntanglementAsymmetryIn11Dimensional,TurkeshiCalabreseDeLuca2025,QuenchingFromSuperfluidToFreeBosonsIn,QuantumMpembaEffectInFreeFermionicMixed,TranslationSymmetryRestorationUnderRandomUnitaryDynamics,EntanglementAsymmetryDynamicsInRandomQuantumCircuits,TuningTheQuantumMpembaEffectInAn,SymmetryBreakingDynamicsInQuantumManyBody,QuantumMpembaEffectWithoutGlobalSymmetries,ExpeditedThermalizationDynamicsInIncommensurateSystems,QuantumMpembaEffectInLocalGaugeSymmetryRestoration,pnf1-r1rm,xu2025observationmodulationquantummpemba,ImaginaryTimeMpembaEffectInQuantumMany,QuantumMpembaEffectInAFourSite,QuantumMpembaEffectInLongRangedU,StarkManyBodyLocalizationInducedQuantumMpemba,QuantumPontusMpembaEffectsInRealAndImaginaryTimeDynamics}. Beyond its conceptual interest, the Mpemba effect carries a direct operational message for quantum state preparation~\cite{g94p-7421}: the fastest route to a target state need not originate from the apparently closest initial condition.

Despite their common relevance to quantum state preparation, quantum magic and quantum Mpemba physics have not yet been directly linked.
Prior work on magic dynamics in random-circuit models found no Mpemba behavior for magic, even though it arises for other quantum resources~\cite{aditya2025mpembaeffectsquantumcomplexity}.
However, conservation laws are known to substantially reshape both the growth and steady-state values of magic~\cite{QuantumMagicDynamicsInRandomCircuits,NonstabilizernessInU1LatticeGaugeTheory,ANonstabilizernessMonotoneFromStabilizernessAsymmetry,iannotti2026nonstabilizernessu1symmetrychaotic}, suggesting that symmetry constraints may qualitatively alter the dynamical generation of nonstabilizer resources.
This raises a central question: can a state with lower initial magic generate nonstabilizer resources faster than one with higher initial magic under symmetry-constrained dynamics, and what microscopic features of the initial state control whether such a crossing occurs?

To answer this question, we study the dynamics of quantum magic, quantified by the second stabilizer R\'enyi entropy $M_2$, in $\mathrm{U(1)}$-symmetric random circuits initialized from tilted product states.
We first derive analytical predictions for the late-time $M_2$ for several classes of tilted initial states, showing that the conserved charge-sector distribution strongly constrains the steady-state magic and produces systematic deviations from the symmetry-free Haar prediction~\cite{Leone2022SRE,Iannotti2025entanglement}.
Turning to the transient dynamics, we uncover an inverse Mpemba effect in magic generation for tilted ferromagnetic states, governed by the interplay between charge conservation, sector-resolved thermalization, and the initial charge distribution.
The charge distribution, however, is not the whole story: tilted N\'eel and tilted domain-wall states share identical charge-sector distributions, yet only the latter exhibits the inverse Mpemba effect, showing that the dynamics also depend on the spatial structure of the initial state within each charge sector.

The nonstabilizerness Mpemba effects extend beyond random circuits.
We further demonstrate them in a nonintegrable quantum spin chain, where the tilted state families exhibiting Mpemba behavior differ from the $\mathrm{U(1)}$ case.
Together with an $\mathrm{SU(2)}$-symmetric circuit example, this shows that magic Mpemba effects are tied neither to Abelian symmetry nor to random-circuit dynamics.
These findings complement recent studies focused on symmetry-constrained steady-state magic and open a route to exploiting symmetry, locality, and conservation laws for the efficient preparation of target many-body resource states, including those relevant to measurement-based quantum computation~\cite{AOneWayQuantumComputer,MeasurementBasedQuantumComputationOnClusterStates,UniversalResourcesForMeasurementBasedQuantumComputation,MeasurementBasedQuantumComputation}.

\begin{figure}[t]
\centering
\includegraphics[width=0.99\columnwidth]{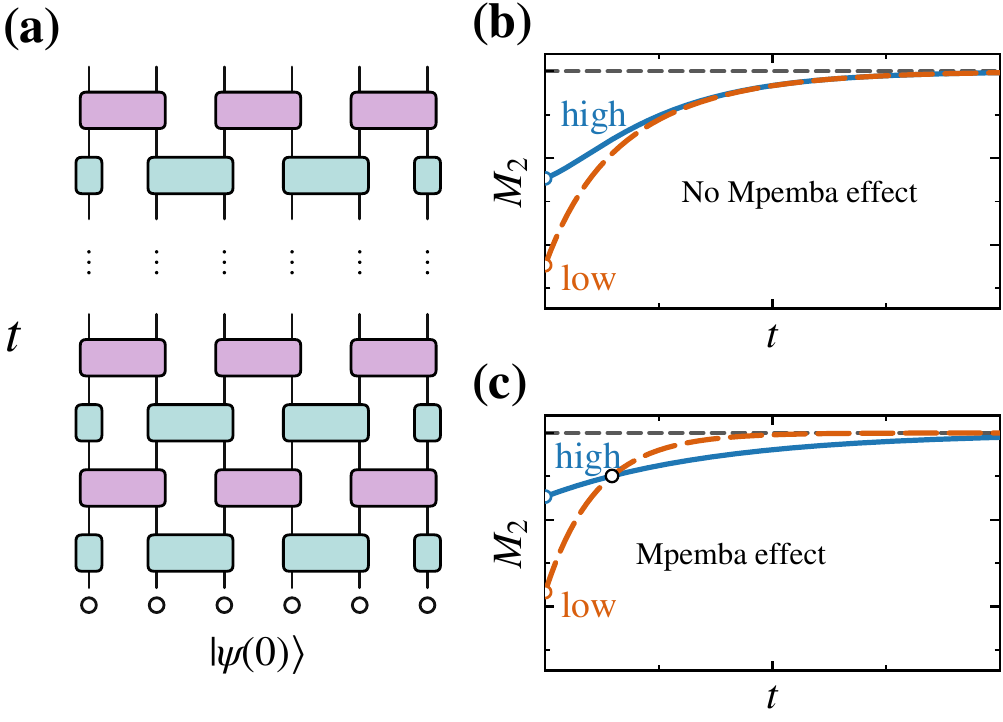}
\caption{(a) Schematic of the $\mathrm{U(1)}$-symmetric brickwork random circuit. 
Each time step consists of two layers of nearest-neighbor two-qubit gates with periodic boundary conditions. 
(b),(c) Schematic illustration of magic dynamics without and with an inverse Mpemba effect, respectively.}
\label{fig:schematic}
\end{figure}

\textbf{Circuit ensemble and magic diagnostic.---}
We study a brickwork random circuit (Fig.\,\ref{fig:schematic}) of nearest-neighbor two-qubit gates conserving the total $\mathrm{U(1)}$ charge $Q=\sum_{j=1}^{N} n_j$, with periodic boundary conditions~\cite{PhysRevX.12.041002,PhysRevLett.129.120604, PhysRevLett.129.200602, PhysRevB.107.014308}.
Each local gate has the charge-block-diagonal form
\begin{align}
U_{j,j+1}
=
e^{i\phi_0}\ket{00}\!\bra{00}
\oplus
U^{(1)}
\oplus
e^{i\phi_2}\ket{11}\!\bra{11},
\label{eq:gate}
\end{align}
with $U^{(1)}$ a $2\times2$ Haar-random unitary on $\mathrm{span}\{\ket{01},\ket{10}\}$ and $\phi_0,\phi_2$ independent random phases. One time step alternates two layers of such gates on even and odd bonds.

Magic can be quantified by several measures, including mana~\cite{Veitch2014Resource}, robustness of magic~\cite{RobustnessOfMagicAndSymmetriesOfThe}, stabilizer rank~\cite{TradingClassicalAndQuantumComputationalResources}, and stabilizer extent~\cite{SimulationOfQuantumCircuitsByLowRankStabilizerDecompositions}; we adopt the stabilizer R\'enyi entropy~\cite{Leone2022SRE}.
For an $N$-qubit pure state $\ket{\psi}$, expand
\begin{align}
    \rho=\ket{\psi}\bra{\psi}
    =
    \frac{1}{2^N}\sum_{P\in\mathcal{P}_N} c_P P,
    \qquad
    c_P=\bra{\psi}P\ket{\psi},
\end{align}
with $\mathcal{P}_N=\{I,X,Y,Z\}^{\otimes N}$. The $\alpha$-th stabilizer R\'enyi entropy is
\begin{align}
    M_\alpha(\ket{\psi})
    =
    \frac{1}{1-\alpha}
    \log_2\!\left[
        \sum_{P\in\mathcal{P}_N}
        \left(
            \frac{|c_P|^2}{2^N}
        \right)^\alpha
    \right]
    -N .
    \label{eq:M_alpha}
\end{align}
We focus on $M_2$, which is a nonstabilizerness monotone for pure states~\cite{StabilizerEntropiesAndNonstabilizernessMonotones,StabilizerEntropiesAreMonotonesForMagicState}
and experimentally accessible via randomized measurements~\cite{PredictingManyPropertiesOfAQuantumSystem,ProbingRenyiEntanglementEntropyViaRandomizedMeasurements}.

In the present setting, we identify that an inverse Mpemba effect occurs when two initial states obey $M_2[\psi_1(0)]<M_2[\psi_2(0)]$, yet evolve under the same circuit ensemble such that there exists a crossing time $t_M$ after which
\begin{align}
    M_2[\psi_1(t)]>M_2[\psi_2(t)],
    \qquad t>t_M .
\end{align}
Thus, the state that is less magic initially becomes more magic at later times.

\textbf{Charge sectors and initial state families.---} 
We use symmetry-breaking tilted product states as initial states, which have been widely used in studies of quantum Mpemba effects associated with symmetry restoration~\cite{EntanglementAsymmetryAsAProbeOfSymmetry,LiuZhangYinZhang2024,TurkeshiCalabreseDeLuca2025,QuantumMpembaEffectsInManyBodyLocalization}. 
Specifically, we consider three state families obtained by applying a uniform Pauli-$Y$ rotation to ferromagnetic, N\'eel, and domain-wall product configurations. 
The tilted ferromagnetic states are
\begin{align}
\ket{\psi(\theta)}_{\rm TFS}
&=
e^{-i\frac{\theta}{2}\sum_{j=1}^{N}Y_j}
\ket{0}^{\otimes N}
\nonumber\\
&=
\bigotimes_{j=1}^{N}
\left(
\cos\frac{\theta}{2}\ket{0}
+
\sin\frac{\theta}{2}\ket{1}
\right),
\label{eq:TFS}
\end{align}
where $\theta$ is the tilt angle. 
The tilted N\'eel states and tilted domain-wall states are defined analogously by replacing the reference configuration $\ket{0}^{\otimes N}$ in Eq.~\eqref{eq:TFS} with
$\ket{01}^{\otimes N/2}$ and 
$\ket{0}^{\otimes N/2}\otimes\ket{1}^{\otimes N/2}$, respectively.

\begin{figure}[t]
\centering
\includegraphics[width=0.99\columnwidth]{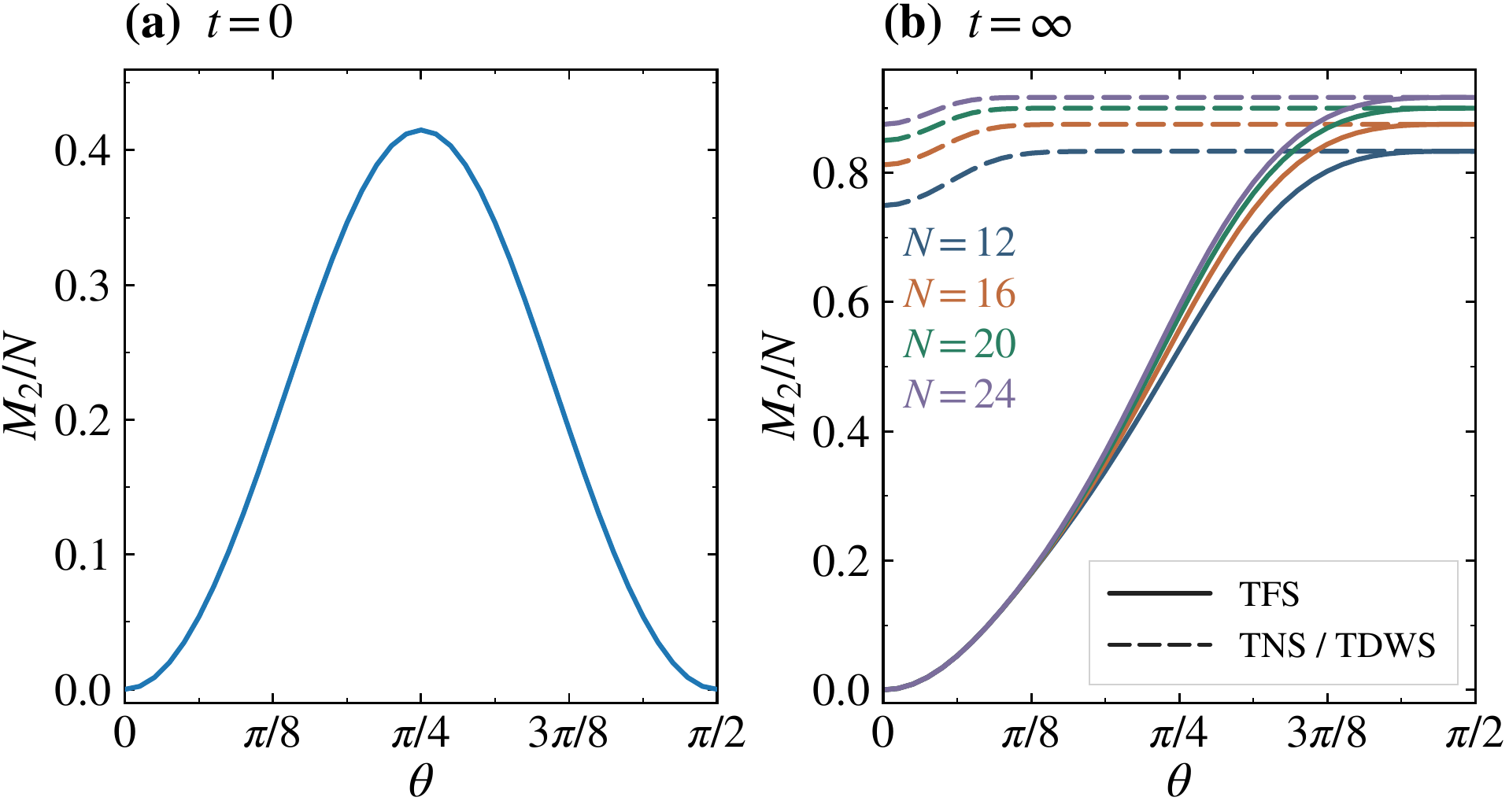}
\caption{Initial and late-time second stabilizer R\'enyi entropy densities. 
(a) Initial value $M_2(0)/N$, which is identical for the families of tilted ferromagnetic states (TFS), tilted N\'eel states (TNS), and tilted domain-wall states (TDWS). 
(b) Late-time global-$\mathrm{U(1)}$ Haar prediction for $M_2^{(\infty)}/N$ for TFS (solid lines) and TNS (dashed lines); TDWS coincides with TNS in this limit. 
Colors indicate system sizes $N=12,16,20,24$.}
\label{fig:long_time_M2}
\end{figure}

All three families have the same initial stabilizer R\'enyi entropy, given as [see the detailed derivation in the Supplemental Material (SM)~\cite{SupplementalMaterials}]
\begin{align}
M_2(\theta)
=
-N\log_2\!\left[
1-\frac{1}{4}\sin^2(2\theta)
\right].
\label{eq:M2_main}
\end{align}
As illustrated in Fig.\,\ref{fig:long_time_M2}(a), at $\theta=0$ and $\theta=\pi/2$, one has $M_2=0$, since the corresponding states are product states in the $Z$ and $X$ bases, respectively, and are therefore stabilizer states.
$M_2$ increases with $\theta$ for $\theta\in[0,\pi/4]$, reaches its maximum at $\theta=\pi/4$, and decreases for $\theta\in[\pi/4,\pi/2]$.

The initial state can be decomposed as
\begin{align}
\ket{\psi(0)}
=
\sum_{q=0}^{N}
\sqrt{p(q)}\,\ket{\psi_q},
\label{eq:charge_decomposition}
\end{align}
where $p(q)$ is the weight of the charge-$q$ sector [$p(q) \geq 0$, $\sum_{q} p(q) = 1$] and $\ket{\psi_q}$ is a normalized state within that sector. 
Because the circuit conserves the total charge $Q$, the weights $p(q)$ are conserved during the evolution, and each $\ket{\psi_q}$ evolves only inside its corresponding charge subspace. 
The charge distribution, therefore, provides a natural, though incomplete, characterization of the ensuing magic dynamics. 
For the tilted-ferromagnetic-state family, increasing $\theta$ over $\theta\in[0,\pi/2]$ shifts weight toward the half-filled sector $q=N/2$~\cite{LiuZhangYinZhang2024}. 
By contrast, the tilted-N\'eel-state and tilted-domain-wall-state families have identical charge distributions, and their dominant sector is always the half-filled sector~\cite{LiuZhangYinZhang2024}, independent of $\theta$ [see analytical expressions for $p(q)$ in the SM~\cite{SupplementalMaterials}]. 
As shown below, however, this conserved distribution alone does not determine the growth of magic.

\textbf{Theoretical $M_{2}$ in the long-time limit.---} We first analyze the long-time limit, where the circuit approaches an effective global $\mathrm{U(1)}$-symmetric random unitary, i.e., a block-diagonal unitary acting within each fixed-charge sector~\cite{brandaoLocalRandomQuantum2016,PhysRevX.15.021022,PRXQuantum.5.040349,liSUdsymmetricRandomUnitaries2025a}. 
Unlike the single-sector case~\cite{iannotti2026nonstabilizernessu1symmetrychaotic}, the states considered here are coherent superpositions of different charge sectors, so $M_2$ is not a weighted sum of fixed-sector values: the Pauli strings in its definition couple different sectors and probe inter-sector coherences. 
In the SM~\cite{SupplementalMaterials}, we derive the analytical late-time value of $M_2$ for all three tilted state families.
The resulting predictions as a function of the tilt angle $\theta$ are shown in Fig.\,\ref{fig:long_time_M2}(b) for several system sizes $N$, with tilted N\'eel states and tilted domain-wall states yielding identical late-time values.

For the tilted ferromagnetic states [Fig.\,\ref{fig:long_time_M2}(b)], the late-time $M_2$ is extensive in $N$ with a slope that increases monotonically with $\theta$ plus a subleading correction that becomes pronounced for $\theta \gtrsim \pi/8$. This follows from the charge-sector decomposition: smaller $\theta$ concentrates the state on low-charge sectors with restricted Hilbert space, suppressing the steady-state magic; larger $\theta$ shifts weight to higher-charge sectors and raises $M_2$.

By contrast, the $\theta$-dependence is much weaker for the tilted N\'eel states and tilted domain-wall states. For the system sizes considered here, once $\theta \gtrsim \pi/8$, the late-time value of $M_2/N$ becomes nearly independent of $\theta$. This weak dependence reflects that, for both states, the charge distribution is already dominated by the half-filled sector across most tilt angles considered.

At $\theta=\pi/2$, all three tilted families share the binomial weights $p(q)=2^{-N}\binom{N}{q}$, and the large-$N$ steady-state $M_2$ is
\begin{align}
M_2^{(\infty)}\!\left(\frac{\pi}{2},N\right)
=
N-2+\frac{3}{(\ln 2)\,2^N}+O(4^{-N}),
\label{eq:M2_infty_pi_over_2}
\end{align}
matching the symmetry-free Haar prediction~\cite{Leone2022SRE,Iannotti2025entanglement} up to exponentially small corrections. No comparably compact closed form is available for general $\theta$; the finite-$N$ expression is derived in the SM~\cite{SupplementalMaterials}.

\begin{figure*}[t]
\centering
\includegraphics[width=1.98\columnwidth]{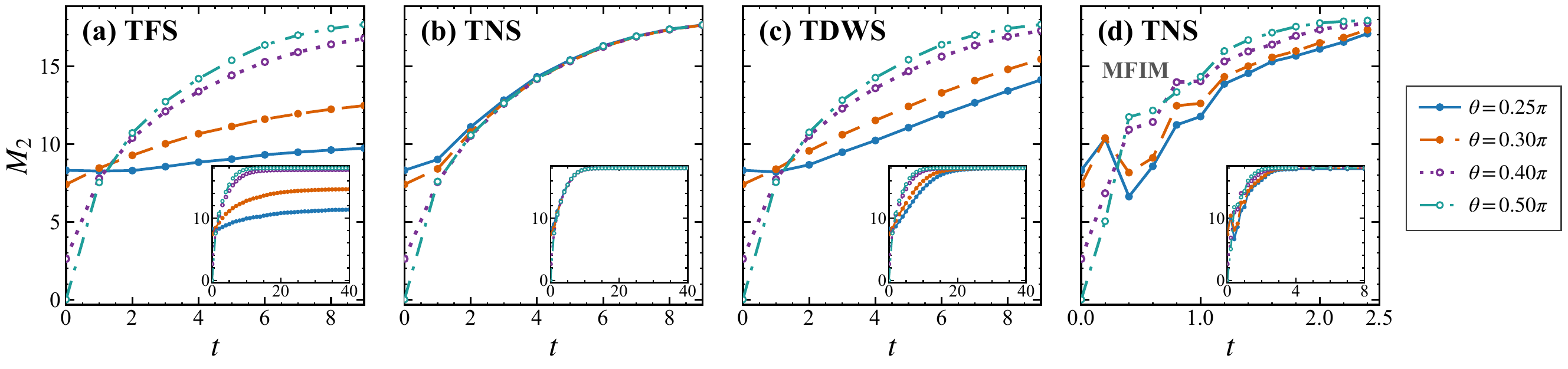}
\caption{Magic dynamics under $\mathrm{U(1)}$-symmetric circuit evolution [(a)--(c)] and mixed-field Ising Hamiltonian (MFIM) evolution [(d)] for $N=20$ and several tilt angles $\theta\in[\pi/4,\pi/2]$. 
Panels (a) and (c), corresponding to tilted ferromagnetic states (TFS) and tilted domain-wall states (TDWS), respectively, exhibit inverse-Mpemba crossings, whereas panel (b), corresponding to tilted N\'eel states (TNS), shows no inverse Mpemba effect. 
(d) In contrast to the $\mathrm{U(1)}$-symmetric circuit dynamics in (b), the same TNS initial states under MFIM evolution exhibit Mpemba-type magic dynamics.}
\label{fig:dynamics}
\end{figure*}

\begin{table}[t]
\caption{Comparison between three classes of initial states: tilted ferromagnetic states (TFS), tilted N\'eel states (TNS), and tilted domain-wall states (TDWS). Here $p(q)$ denotes the probability that the state lies in the total-charge-$q$ sector.}
\begin{ruledtabular}
\begin{tabular}{lccc}
Family & Same $M_2(0)$ & Same $p(q)$ & Inverse Mpemba \\
\hline
TFS  & \cmark & \xmark   & \cmark \\
TNS  & \cmark & \cmark ($=$ TDWS) & \xmark \\
TDWS & \cmark & \cmark ($=$ TNS)  & \cmark
\end{tabular}
\end{ruledtabular}
\label{tab:summary}
\end{table}

\textbf{Mpemba effect in early-time dynamics.---} Using recent methods for computing $M_2$~\cite{xiao2026exponentiallyacceleratedsamplingpauli,huang2026fastexactapproachstabilizer,Sierant2026computingquantum}, we simulate the early-time magic dynamics for $\theta \in [\pi/4,\pi/2]$, where the initial $M_2$ decreases monotonically with $\theta$.

For the tilted ferromagnetic states, as shown in Fig.\,\ref{fig:long_time_M2}(b), the late-time value of $M_2$ increases monotonically with $\theta$ throughout the regime $\theta \in [\pi/4,\pi/2]$. Combined with the corresponding monotonic decrease of the initial $M_2$ in Fig.\,\ref{fig:long_time_M2}(a), this immediately implies that, for any two tilt angles satisfying $\theta_1 < \theta_2$, the two trajectories must cross at some intermediate time $t_M$, thereby giving rise to the quantum Mpemba effect. The physical origin of this behavior can be understood from the perspective of quantum thermalization in the presence of a conserved charge. As $\theta$ increases, the dominant charge sector moves closer to half filling, where the Hilbert-space dimension is larger. The enlarged effective Hilbert space facilitates faster thermalization and more rapid magic generation, so that the state with smaller initial magic can nevertheless relax more quickly.

More interestingly, magic generation is not determined solely by the charge distribution $p(q)$. We therefore compare the magic dynamics of the tilted N\'eel states and tilted domain-wall states, which share the same $p(q)$ at fixed $\theta$. Since the late-time values of $M_2$ are nearly flat throughout the regime $\theta \in [\pi/4,\pi/2]$, the presence or absence of the quantum Mpemba effect is controlled primarily by the early-time dynamics.

For the tilted N\'eel states, by contrast, no quantum Mpemba effect is observed, as shown in Fig.\,\ref{fig:dynamics}(b). Throughout the evolution, the state with the larger initial magic retains the larger $M_2$ until both trajectories eventually converge to the same long-time value. Since the dominant charge sector remains half-filled throughout, the thermalization speed is essentially insensitive to $\theta$. As a result, the state with the larger initial magic approaches the common late-time value from above without crossing. Indeed, the magic dynamics for different $\theta$ nearly collapse onto a single curve already after the first time step.

For the tilted domain-wall states, however, the quantum Mpemba effect persists, as shown in Fig.\,\ref{fig:dynamics}(c), in close analogy with the tilted ferromagnetic state case. This is more nontrivial because the dominant charge sector is again half-filled for all $\theta$. The key difference lies in the local structure of the initial state: the domain-wall configuration contains extended locally ferromagnetic regions on which the $\mathrm{U(1)}$-symmetric gates act trivially or only weakly, thereby slowing the early-time thermalization, especially at small $\theta$. In the small-$\theta$ regime, only the gate acting across the domain wall initially contributes significantly to the dynamics, so relaxation proceeds through the gradual spreading of the domain wall over a timescale that grows approximately linearly with system size. Consequently, for the same $\theta$, magic generation from the tilted domain-wall states is much slower than that from the tilted N\'eel states, as is evident from comparing Fig.\,\ref{fig:dynamics}(b) and Fig.\,\ref{fig:dynamics}(c). In particular, when $\theta$ is small, the approach to the late-time value is delayed by an additional timescale of order $O(L)$. More importantly, despite this slow global relaxation, the quantum Mpemba effect still emerges at early times, driven by the same local mechanism operative in the tilted ferromagnetic states.

We summarize these findings for the different tilted product state families in Table~\ref{tab:summary}.
Therefore, although the late-time value of $M_2$ is determined solely by the initial charge distribution $p(q)$, the early-time dynamics, and hence the presence or absence of the quantum Mpemba effect, are controlled not only by $p(q)$ but also by the structure within each charge sector.

\textbf{Nonstabilizerness Mpemba effect in Hamiltonian dynamics.---}
To test whether the magic Mpemba effect requires an exact internal symmetry, we evolve the same tilted product states under the nonintegrable mixed-field Ising model (MFIM)
\begin{equation}
H=-\sum_i Z_iZ_{i+1}-h_x\sum_i X_i-h_z\sum_i Z_i ,
\end{equation}
with periodic boundary conditions and $(h_x,h_z)=(-1.05,0.5)$, which has no $\mathrm{U(1)}$ charge conservation.
The tilted N\'eel family gives the sharpest contrast: it shows no magic Mpemba effect under the $\mathrm{U(1)}$ circuit, yet a clear one under the MFIM evolution [Fig.\,\ref{fig:dynamics}(d)].
States with larger $\theta \in (\pi/4,\pi/2]$ start with lower $M_2$, overtake the $\theta=\pi/4$ reference at early times, and approach the high-$M_2$ plateau earlier [Fig.\,\ref{fig:dynamics}(d)].

This contrast with the $\mathrm{U(1)}$-symmetric circuit follows from the initial-state energy.
The MFIM Hamiltonian is traceless, so $e_\infty=2^{-N}\mathrm{Tr}\,H/N=0$ marks the spectrum center.
For the tilted N\'eel family the field terms cancel between sublattices and $\langle Z_iZ_{i+1}\rangle=-\cos^2\theta$ gives
\begin{equation}
{\langle H\rangle_{\rm TNS}}/{N}=\cos^2\theta ,
\end{equation}
so increasing $\theta$ in $(\pi/4,\pi/2]$ moves the state toward the spectrum center.
Energy now plays the role of the conserved quantity that $\mathrm{U(1)}$ charge plays in the circuit~\cite{Deutsch1991QuantumStatisticalMechanics,Srednicki1994ChaosQuantumThermalization,Rigol2008ThermalizationMechanism,DAlessio2016QuantumChaosETH,KimHuse2013BallisticSpreading}.
In nonintegrable spin chains, energy diffuses with a diffusion constant that depends on energy density~\cite{ZanociSwingle2021TemperatureDependentEnergyDiffusion}.
Different $\theta$ therefore correspond to different effective relaxation rates.
Larger-$\theta$ TNS states sit nearer infinite temperature and relax faster, accounting for the observed crossing.
The tilted domain-wall family behaves the same way; the tilted ferromagnetic family shows no Mpemba effect (see SM~\cite{SupplementalMaterials}).

\textbf{Discussion and outlook.---}
In this work, we study the dynamical generation of second stabilizer R\'enyi entropy $M_2$ in $\mathrm{U(1)}$-symmetric random circuits initialized from several classes of tilted product states.
We derived analytical predictions for the late-time $M_2$, showing that the conserved charge-sector distribution strongly constrains the steady-state magic and produces systematic deviations from the symmetry-free Haar value.
The early-time dynamics, however, is not controlled by the charge distribution alone. 
By comparing tilted N\'eel and tilted domain-wall states with identical charge-sector probabilities, we identified the spatial structure of the initial state within each charge sector as an additional control parameter governing magic generation.
The same mechanism extends to $\mathrm{SU(2)}$-symmetric random circuits (see SM~\cite{SupplementalMaterials}) and to nonintegrable mixed-field Ising Hamiltonian dynamics, where energy plays the role of the conserved quantity analogous to U(1) charge and induces an effective sector structure on its own~\cite{PhysRevX.14.031014,cui2025randomunitarieshamiltoniandynamics,mao2025randomunitariesconserveenergy}.
These results establish that magic Mpemba effects are tied neither to Abelian symmetry nor to random-circuit dynamics.

Several directions remain open.
In integrable and many-body localized systems~\cite{PhysRevB.75.155111, basko2006metal, annurev:/content/journals/10.1146/annurev-conmatphys-031214-014726, annurev:/content/journals/10.1146/annurev-conmatphys-031214-014701, PhysRevLett.117.040601, https://doi.org/10.1002/andp.201700169, PhysRevLett.130.120403,PhysRevLett.121.206601,Sierant_2025,PhysRevLett.119.206602}, conventional thermalization fails and qualitatively different mechanisms may emerge for magic generation.
The interplay between non-Abelian symmetry and magic dynamics, which our $\mathrm{SU(2)}$ example only begins to probe (see SM~\cite{SupplementalMaterials}), deserves a more systematic investigation.
The mechanisms identified here also suggest analogous anomalous early-time dynamics for other quantum resources, including coherence~\cite{PhysRevLett.113.140401,RevModPhys.89.041003,PhysRevResearch.2.023298,aditya2026coherencedynamicsquantummanybody}, imaginarity~\cite{chen2025imaginarity,xu2024quantifying,PhysRevA.111.032425,Wu_2025}, and non-Gaussianity~\cite{PhysRevResearch.6.023176,PhysRevLett.123.080503}.
From a practical perspective, these results indicate that Mpemba-type acceleration may serve as a useful guiding principle for the efficient preparation of target many-body resource states.

\textbf{Acknowledgments}--- We acknowledge helpful discussions with Shi-Xin Zhang and Yi-Mu Bao.
Z.X. is supported by the Princeton Quantum Initiative Fellowship.
S.L. was supported by the Gordon and Betty Moore Foundation through Grant No. GBMF8685 towards the Princeton theory program, the Gordon and Betty Moore Foundation’s EPiQS Initiative (Grant No. GBMF11070), the Global Collaborative
Network Grant at Princeton University, the Simons Investigator Grant No. 404513, the Princeton Global
Network, the NSF-MERSEC (Grant No. MERSEC DMR 2011750), the Simons Collaboration on New Frontiers in Superconductivity (Grant No. SFI-MPS-NFS-00006741-01 and No. SFI-MPS-NFS-00006741-06), the Princeton Catalysis
Initiative, the Schmidt Foundation at the Princeton University,
European Research Council (ERC) under the European Union’s Horizon 2020 research and innovation program (Grant Agreement No. 101020833), the National Science Foundation through the AI Research Institutes program
Award No. DMR-2433348. H.K.Z.  was supported by the Postdoctoral Fellowship Program and China Postdoctoral Science Foundation (No. BX20250169) and Beijing Natural Science Foundation (No. 1264076).

%

\clearpage
\newpage
\widetext

\begin{center}
\textbf{\large Supplemental Material for ``Nonstabilizerness Mpemba Effects''}
\end{center}

\renewcommand{\thefigure}{S\arabic{figure}}
\renewcommand{\theHfigure}{S\arabic{figure}}
\setcounter{figure}{0}
\renewcommand{\theequation}{S\arabic{equation}}
\renewcommand{\theHequation}{S\arabic{equation}}
\setcounter{equation}{0}
\renewcommand{\thesection}{\Roman{section}}
\renewcommand{\theHsection}{S\arabic{section}}
\setcounter{section}{0}
\setcounter{secnumdepth}{4}

\addtocontents{toc}{\protect\setcounter{tocdepth}{0}}
{
\tableofcontents
}

\section{\texorpdfstring{$\mathrm{SU(2)}$}{SU(2)}-Symmetric Circuit}

To test whether the same mechanism survives beyond the Abelian setting, we also
studied brickwork random circuits built from nearest-neighbor two-qubit gates
that commute with simultaneous spin rotations on each bond. For two spin-$1/2$
degrees of freedom, the local Hilbert space decomposes as
\begin{align}
\frac{1}{2}\otimes\frac{1}{2}=0\oplus 1,
\end{align}
namely into a singlet and a triplet sector. Consequently, the most general
two-qubit $\mathrm{SU(2)}$-symmetric gate has the form
\begin{align}
U_{j,j+1}^{\mathrm{SU(2)}}
=
e^{i\phi_{j,j+1}^{(0)}}P_{j,j+1}^{(0)}
+
e^{i\phi_{j,j+1}^{(1)}}P_{j,j+1}^{(1)},
\label{eq:sm_su2_gate}
\end{align}
where
\begin{align}
P_{j,j+1}^{(0)}=\ket{\Psi^-_{j,j+1}}\!\bra{\Psi^-_{j,j+1}},
\qquad
P_{j,j+1}^{(1)}=I_{j,j+1}-P_{j,j+1}^{(0)},
\end{align}
with
\begin{align}
\ket{\Psi^-_{j,j+1}}
=
\frac{1}{\sqrt{2}}
\left(
\ket{01}_{j,j+1}-\ket{10}_{j,j+1}
\right).
\end{align}
In the numerics, the phases $\phi_{j,j+1}^{(0)}$ and $\phi_{j,j+1}^{(1)}$ are
drawn independently and uniformly from $[0,2\pi)$ for every gate.

As initial states we use a staggered tilted ferromagnetic family, obtained
by rotating the fully polarized state by opposite angles on the two
sublattices:
\begin{align}
\ket{\psi(\theta)}_{\rm sTFS}
&=
\exp\!\left[
-i\frac{\theta}{2}\sum_{j=1}^{N}(-1)^{j-1}Y_j
\right]
\ket{0}^{\otimes N}
\nonumber\\
&=
\bigotimes_{j=1}^{N}
\left(
\cos\frac{\theta}{2}\ket{0}
+
(-1)^{j-1}\sin\frac{\theta}{2}\ket{1}
\right).
\label{eq:sm_su2_stfs}
\end{align}
Up to the alternating sign in the $\ket{1}$ amplitude, each site carries the
same local polar angle as in Eq.~\eqref{eq:TFS}, so the initial product-state
magic remains controlled by the single-qubit tilt angle $\theta$.
Crucially, however, $\ket{\psi(\theta)}_{\rm sTFS}$ is not obtained from
$\ket{0}^{\otimes N}$ by a \emph{global} spin rotation, and is therefore not
confined to the maximal-spin irrep $S=N/2$; it instead admits a nontrivial
decomposition over total-spin sectors and the associated multiplicity spaces,
with $\theta$-dependent weights.

\begin{figure*}[t]
\centering
\includegraphics[width=0.65\columnwidth]{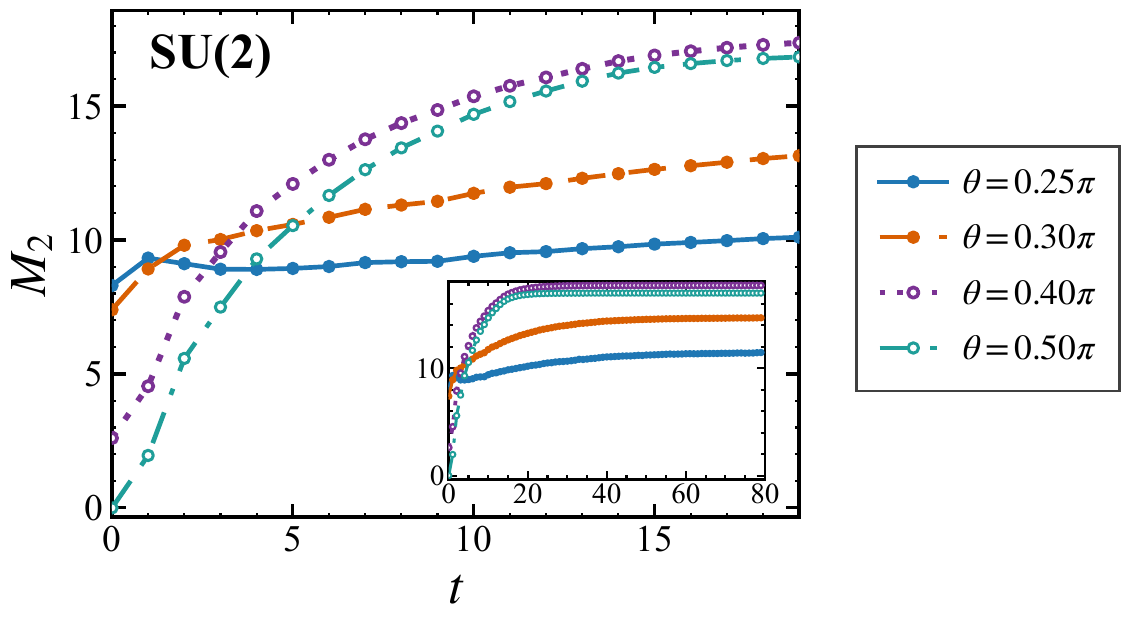}
\caption{Magic dynamics under the same brickwork $\mathrm{SU(2)}$-symmetric random
circuit for the staggered tilted ferromagnetic states
\eqref{eq:sm_su2_stfs} at $N=20$. The main panel shows early and intermediate
times, while the inset extends the evolution to $t=80$. Lower-initial-magic
states with $\theta=0.40\pi$ and $\theta=0.50\pi$ overtake the more magical
$\theta=0.25\pi$ and $\theta=0.30\pi$ trajectories, demonstrating
inverse-Mpemba-type crossings also in the non-Abelian setting.}
\label{fig:su2_dynamics}
\end{figure*}

Figure~\ref{fig:su2_dynamics} shows that inverse-Mpemba-type behavior persists in this $\mathrm{SU(2)}$-symmetric circuit.
The late-time ordering is nontrivial: among the angles shown, the $\theta=0.40\pi$ curve reaches the largest long-time value, while the $\theta=0.50\pi$ curve also overtakes the initially more magic states at $\theta=0.25\pi$ and $\theta=0.30\pi$.
This $\mathrm{SU(2)}$ example demonstrates that anomalous magic generation is not restricted to Abelian symmetry constraints. We note, however, that no crossing occurs between the curves with $\theta=0.40\pi$ and $\theta=0.50\pi$, a feature whose underlying mechanism deserves further investigation.

\section{Mixed-Field Ising Model}

\begin{figure*}[t]
\centering
\includegraphics[width=0.78\textwidth]{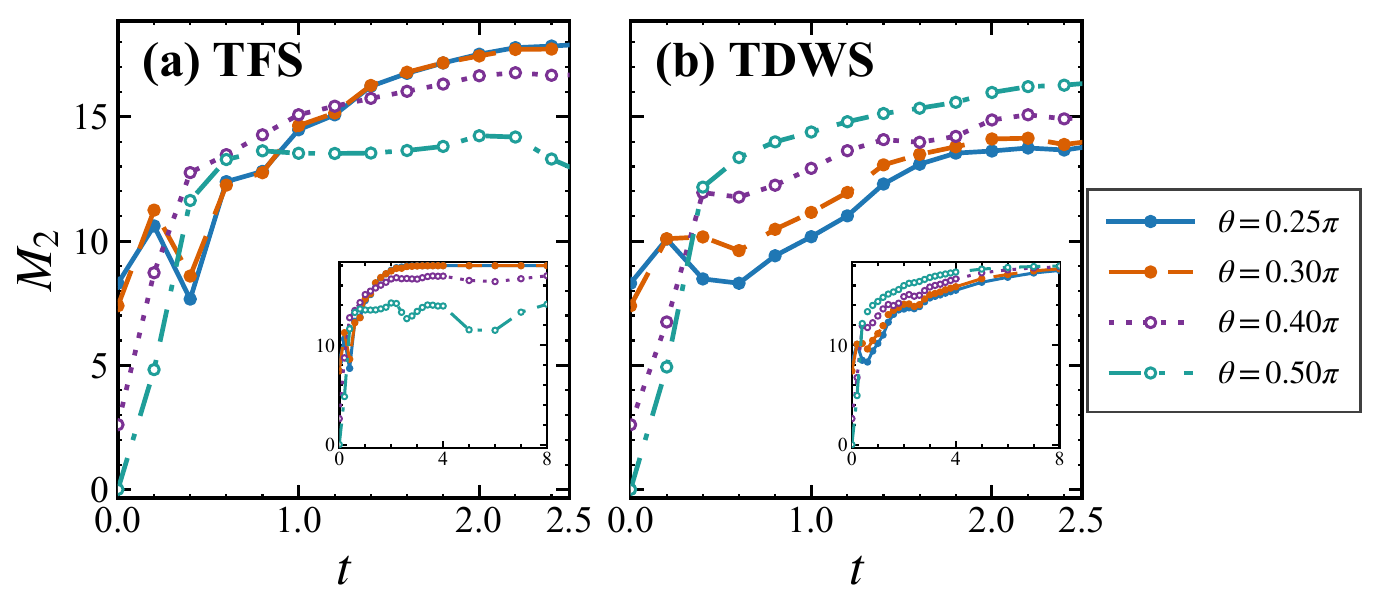}
\caption{Mixed-field Ising-model magic dynamics for the tilted ferromagnetic
states and tilted domain-wall states at $N=20$, with
$(h_x,h_z)=(-1.05,0.5)$. The main panels show the early-time window
$0\le t\le 2.5$, and the insets extend the same trajectories to $t=8$. (a) The
tilted ferromagnetic family shows transient oscillatory crossings. (b) The
tilted domain-wall family shows the same mixed-field Ising-model Mpemba
behavior as the tilted N\'eel family in Fig.\,\ref{fig:dynamics}(d).}
\label{fig:sm_mfim_tfs_tdws}
\end{figure*}

Figure~\ref{fig:sm_mfim_tfs_tdws} shows the tilted ferromagnetic and tilted
domain-wall families under the same mixed-field Ising-model evolution as the main text, with
$(h_x,h_z)=(-1.05,0.5)$.

The tilted domain-wall family follows the same energy-based mechanism as the
tilted N\'eel family. The relevant energy densities are
\begin{align}
\langle H\rangle_{\rm TNS}/N&=\cos^2\theta,\\
\langle H\rangle_{\rm TDWS}/N&=-(1-4/N)\cos^2\theta.
\end{align}
Both move toward the spectrum center $e=0$ as $\theta$ increases, and as in the
tilted N\'eel case, larger-$\theta$ tilted domain-wall states relax faster and
overtake the smaller-$\theta$ trajectories.

The tilted ferromagnetic family behaves differently. Its energy density is
\begin{equation}
\langle H\rangle_{\rm TFS}/N=-\cos^2\theta+1.05\sin\theta-0.5\cos\theta .
\end{equation}
This places the large-$\theta$ states near the upper spectral edge.
We numerically find that the overlap of the tilted ferromagnetic states with
the highest-energy eigenstate $|E_{\max}\rangle$ is given by
\begin{equation}
|\langle E_{\max}|\psi_{\rm TFS}\rangle|^2
=6.44\times10^{-4},\;4.54\times10^{-3},\;6.26\times10^{-2},\;2.00\times10^{-1}
\end{equation}
for $\theta/\pi=0.25,0.30,0.40,0.50$, respectively. 
For $\theta/\pi=0.40,0.50$, the overlap is substantial, resulting in coherent
oscillation. The tilted ferromagnetic family shows no robust magic Mpemba
effect.

\section{Steady-State Second Stabilizer R\'enyi Entropy Under Global \texorpdfstring{$\mathrm{U(1)}$}{U(1)} Haar Scrambling}

We assume that the $\mathrm{U(1)}$-symmetric evolution in sufficiently long time acts as independent Haar-random unitaries on each fixed-charge sector~\cite{brandaoLocalRandomQuantum2016,PhysRevX.15.021022,PRXQuantum.5.040349,liSUdsymmetricRandomUnitaries2025a}, and compute the resulting late-time second stabilizer R\'enyi entropy. The result is an exact finite-$N$ formula expressed in terms of the sector weights $p(q)$. 
The steady-state values for the tilted ferromagnetic, tilted N\'eel, and tilted domain-wall initial states are then obtained by substituting the corresponding charge distributions $p(q)$.

\subsection{Setup}

We use $N$ to denote the number of qubits and $\mathcal H_q$ to denote the fixed-charge subspace with
\begin{align}
D_q=\dim \mathcal H_q=\binom{N}{q},\qquad q=0,1,\dots,N.
\label{eq:sm_sector_dimension}
\end{align}
As in the main text, $q$ is the Hamming weight of the computational basis state. For a general initial state,
\begin{align}
\ket{\psi(0)}=\sum_{q=0}^{N}\sqrt{p(q)}\,\ket{\psi_q(0)},\qquad \braket{\psi_q(0)|\psi_q(0)}=1,
\label{eq:sm_initial_sector_decomp}
\end{align}
the global-$\mathrm{U(1)}$ Haar late-time ansatz is
\begin{align}
\ket{\Psi_\infty}=\sum_{q=0}^{N}\sqrt{p(q)}\,\ket{\phi_q},\qquad \ket{\phi_q}=U_q\ket{\psi_q(0)},
\label{eq:sm_global_haar_state}
\end{align}
where each $U_q\in U(\mathcal H_q)$ is an independent Haar-random unitary. The
weights $p(q)$, fixed by charge conservation, persist; the Haar average
erases all intra-sector information.

For $\alpha=2$, the quantity of interest is the stabilizer purity
\begin{align}
\Xi_2(\ket{\Psi}):=\frac{1}{2^N}\sum_{P\in\mathcal P_N}\left|\bra{\Psi}P\ket{\Psi}\right|^4,
\label{eq:sm_xi2_def}
\end{align}
with $\mathcal P_N=\{I,X,Y,Z\}^{\otimes N}$. The second stabilizer R\'enyi entropy is then
\begin{align}
M_2(\ket{\Psi})=-\log_2 \Xi_2(\ket{\Psi}).
\label{eq:sm_m2_from_xi2}
\end{align}
We therefore need the Haar average of $\Xi_2$.

We label computational basis states by bitstrings $\mathbf x=(x_1,\dots,x_N)\in\{0,1\}^N$ and write
\begin{align}
\ket{\Psi}=\sum_{\mathbf x}\Psi_{\mathbf x}\ket{\mathbf x},\qquad |\mathbf x|=\sum_{j=1}^{N}x_j.
\end{align}
Represent a Pauli string by two binary strings $\mathbf a,\mathbf b\in\{0,1\}^N$:
\begin{align}
P_{\mathbf a,\mathbf b}:=i^{\mathbf a\cdot\mathbf b}\,X(\mathbf a)Z(\mathbf b),
\label{eq:sm_binary_pauli}
\end{align}
where
\begin{align}
X(\mathbf a):=\bigotimes_{j=1}^{N}X_j^{a_j},\qquad Z(\mathbf b):=\bigotimes_{j=1}^{N}Z_j^{b_j}.
\end{align}
The overall phase $i^{\mathbf a\cdot\mathbf b}$ drops out of Eq.~\eqref{eq:sm_xi2_def}, so only the bit-flip pattern $\mathbf a$ and the phase pattern $\mathbf b$ matter.

Using
\begin{align}
P_{\mathbf a,\mathbf b}\ket{\mathbf x}=i^{\mathbf a\cdot\mathbf b}(-1)^{\mathbf b\cdot\mathbf x}\ket{\mathbf x\oplus\mathbf a},
\end{align}
one finds
\begin{align}
\bra{\Psi}P_{\mathbf a,\mathbf b}\ket{\Psi}=\sum_{\mathbf x}(-1)^{\mathbf b\cdot\mathbf x}\Psi_{\mathbf x}^{\ast}\Psi_{\mathbf x\oplus\mathbf a},
\label{eq:sm_pauli_expectation_binary}
\end{align}
where $\oplus$ denotes bitwise addition modulo two. Expanding the fourth power and summing over all phase labels $\mathbf b$ gives
\begin{align}
\sum_{\mathbf b}
\left|
\bra{\Psi}P_{\mathbf a,\mathbf b}\ket{\Psi}
\right|^4
&=
\sum_{\mathbf b}
\sum_{\mathbf x,\mathbf y,\mathbf z,\mathbf w}
(-1)^{\mathbf b\cdot(\mathbf x\oplus\mathbf y\oplus\mathbf z\oplus\mathbf w)}
\Psi_{\mathbf x}^{\ast}\Psi_{\mathbf x\oplus\mathbf a}
\Psi_{\mathbf y}^{\ast}\Psi_{\mathbf y\oplus\mathbf a}
\Psi_{\mathbf z}^{\ast}\Psi_{\mathbf z\oplus\mathbf a}
\Psi_{\mathbf w}^{\ast}\Psi_{\mathbf w\oplus\mathbf a}
\nonumber\\
&=
2^N
\sum_{\mathbf x,\mathbf y,\mathbf z}
\Psi_{\mathbf x}^{\ast}\Psi_{\mathbf x\oplus\mathbf a}
\Psi_{\mathbf y}^{\ast}\Psi_{\mathbf y\oplus\mathbf a}
\Psi_{\mathbf z}^{\ast}\Psi_{\mathbf z\oplus\mathbf a}
\Psi_{\mathbf x\oplus\mathbf y\oplus\mathbf z}^{\ast}
\Psi_{\mathbf x\oplus\mathbf y\oplus\mathbf z\oplus\mathbf a},
\label{eq:sm_pauli_fourth_moment_binary}
\end{align}
where in the second line we used the standard identity
\begin{align}
\sum_{\mathbf b}(-1)^{\mathbf b\cdot \mathbf u}=2^N \delta_{\mathbf u,\mathbf 0}.
\end{align}
The $\mathbf b$ sum has been carried out, so
Eq.~\eqref{eq:sm_pauli_fourth_moment_binary} depends on the Pauli label only
through the bit-flip pattern $\mathbf a$. This is the starting point for the
late-time Haar average.

For the state \eqref{eq:sm_global_haar_state}, since $\ket{\phi_q}\in\mathcal{H}_q$, we expand it as
$
\ket{\phi_q}=\sum_{|\mathbf x|=q}\phi_{q,\mathbf x}\ket{\mathbf x},
$ where the complex amplitudes $\phi_{q,\mathbf x}$ are normalized within sector $q$, 
so that the amplitudes in the computational basis can be written as
\begin{align}
\Psi_{\mathbf x}=\sqrt{p(|\mathbf x|)}\,\phi_{|\mathbf x|,\mathbf x},
\label{eq:sm_global_haar_state_Phix}
\end{align}
and the Haar average factorizes sector by sector.

The only sector-wise ingredient we need is the exact Haar moment formula. For a normalized Haar-random vector in a $D_q$-dimensional Hilbert space,
\begin{align}
\overline{
\phi_{q,i_1}^{\ast}\cdots
\phi_{q,i_m}^{\ast}
\phi_{q,j_1}\cdots
\phi_{q,j_m}
}
=
\frac{1}{(D_q)_m}
\sum_{\pi\in S_m}
\prod_{\alpha=1}^{m}
\delta_{i_\alpha,j_{\pi(\alpha)}},
\label{eq:sm_sector_haar_moment}
\end{align}
where
\begin{align}
(D)_m:=D(D+1)\cdots(D+m-1)
\label{eq:sm_rising_factorial}
\end{align}
is the rising factorial, and $S_m$ denotes the symmetric group of permutations of $m$ elements.

To package the sector dependence compactly, define
\begin{align}
\Gamma_{q_1q_2q_3q_4}[p]:=\prod_{a=0}^{N}\frac{p(a)^{\mu_a(q_1,q_2,q_3,q_4)}}{(D_a)_{\mu_a(q_1,q_2,q_3,q_4)}},
\label{eq:sm_gamma_def}
\end{align}
where $\mu_a(q_1,q_2,q_3,q_4)$ is the multiplicity of the sector label $a$ in the multiset $\{q_1,q_2,q_3,q_4\}$. For example,
\begin{align}
\Gamma_{qqrr}[p]
=
\begin{cases}
\dfrac{p(q)^2p(r)^2}{(D_q)_2(D_r)_2}, & q\neq r,\\[0.8em]
\dfrac{p(q)^4}{(D_q)_4}, & q=r.
\end{cases}
\label{eq:sm_gamma_examples}
\end{align}

\subsection{Sector selection rule and $h\ge 1$}

Substituting Eqs.~\eqref{eq:sm_global_haar_state} and \eqref{eq:sm_global_haar_state_Phix} into Eq.~\eqref{eq:sm_pauli_fourth_moment_binary}, together with
\begin{align}
\mathbf w:=\mathbf x\oplus\mathbf y\oplus\mathbf z,
\end{align}
the averaged monomial reads
\begin{align}
&\overline{
\Psi_{\mathbf x}^{\ast}\Psi_{\mathbf x\oplus\mathbf a}
\Psi_{\mathbf y}^{\ast}\Psi_{\mathbf y\oplus\mathbf a}
\Psi_{\mathbf z}^{\ast}\Psi_{\mathbf z\oplus\mathbf a}
\Psi_{\mathbf w}^{\ast}\Psi_{\mathbf w\oplus\mathbf a}
}
\nonumber\\
&=
\left[
\prod_{\ell=1}^{4}\sqrt{p(q_\ell)p(r_\ell)}
\right]
\overline{
\phi_{q_1,\mathbf x}^{\ast}\phi_{r_1,\mathbf x\oplus\mathbf a}
\phi_{q_2,\mathbf y}^{\ast}\phi_{r_2,\mathbf y\oplus\mathbf a}
\phi_{q_3,\mathbf z}^{\ast}\phi_{r_3,\mathbf z\oplus\mathbf a}
\phi_{q_4,\mathbf w}^{\ast}\phi_{r_4,\mathbf w\oplus\mathbf a}
},
\label{eq:sm_basic_monomial_sectorized}
\end{align}
where we introduced the charge labels
\begin{align}
q_1=|\mathbf x|,\quad q_2=|\mathbf y|,\quad q_3=|\mathbf z|,\quad q_4=|\mathbf w|,
\label{eq:sm_q_labels}
\end{align}
and
\begin{align}
r_1=|\mathbf x\oplus\mathbf a|,\quad r_2=|\mathbf y\oplus\mathbf a|,\quad r_3=|\mathbf z\oplus\mathbf a|,\quad r_4=|\mathbf w\oplus\mathbf a|.
\label{eq:sm_r_labels}
\end{align}

Because the random vectors from different charge sectors are independent, the
Haar average in Eq.~\eqref{eq:sm_basic_monomial_sectorized} vanishes unless the
multiset of starred sector labels coincides with the multiset of unstarred
sector labels:
\begin{align}
\{q_1,q_2,q_3,q_4\}_{\mathrm{multiset}}=\{r_1,r_2,r_3,r_4\}_{\mathrm{multiset}}.
\label{eq:sm_multiset_constraint}
\end{align}
When this selection rule holds, together with Eq.~\eqref{eq:sm_sector_haar_moment}, the sector dependence factors into
$\Gamma_{q_1q_2q_3q_4}[p]$~\eqref{eq:sm_gamma_def}, and what remains is a counting problem: how many
bitstrings realize the required charges before and after the flip $\mathbf a$.

Sort Eq.~\eqref{eq:sm_pauli_fourth_moment_binary} by the Hamming weight $h:=|\mathbf a|$. The late-time average depends only on $h$, not on the flip positions, and there are $\binom{N}{h}$ strings $\mathbf a$ at each $h$. The cases $h\ge 1$ and $h=0$ have different coincidence structures, treated in turn.

Fix a representative $\mathbf a$ with $|\mathbf a|=h\ge 1$. For a given computational basis configuration $\mathbf x$, suppose that exactly $k$ of the $h$ flipped sites are occupied in $\mathbf x$. Then the remaining $q-k$ occupied sites must lie among the $N-h$ untouched positions, and after applying the flip pattern $\mathbf a$ the final charge becomes $r=q+h-2k$.
Therefore the number of bitstrings $\mathbf x$ satisfying simultaneously $|\mathbf x|=q$ and $|\mathbf x\oplus \mathbf a|=r$ is
\begin{align}
O_{qr}^{(h)}
&=\sum_{k=0}^{h}\binom{h}{k}\binom{N-h}{q-k}\,\delta_{r,\;q+h-2k}
\nonumber\\
&=[u^qv^r]\,(1+uv)^{N-h}(u+v)^h,
\label{eq:sm_O_def}
\end{align}
where $[u^q v^r]$ denotes the coefficient of $u^q v^r$.

For $h\ge 1$, $\mathbf x$ and $\mathbf x\oplus\mathbf a$ are distinct, so inserting Eq.~\eqref{eq:sm_sector_haar_moment} into Eq.~\eqref{eq:sm_pauli_fourth_moment_binary} leaves three inequivalent pairings of the four amplitudes (the factor of $3$ below), plus a sector-coincidence contribution when two ordered pairs $(\mathbf x,\mathbf x\oplus\mathbf a)$ and $(\mathbf y,\mathbf y\oplus\mathbf a)$ share sector labels.

It is convenient to introduce
\begin{align}
\mathcal A_N^{(h)}[p]&:=2^N\sum_{q,r,m,n=0}^{N}O_{qr}^{(h)}O_{mn}^{(h)}\Gamma_{qrmn}[p],
\label{eq:sm_Ah_def}
\\
\mathcal B_N^{(h)}[p]&:=2^N\sum_{q,r=0}^{N}O_{qr}^{(h)}\Gamma_{qqrr}[p].
\label{eq:sm_Bh_def}
\end{align}
The full $h\ge 1$ contribution is
\begin{align}
\mathcal F_N^{(h)}[p]=3\,\mathcal A_N^{(h)}[p]+6\,\mathcal B_N^{(h)}[p].
\label{eq:sm_F_hpos_final}
\end{align}
Here $\mathcal A_N^{(h)}$ counts the four sector labels independently, while $\mathcal B_N^{(h)}$ collects the additional weight from repeated assignments of the form $(q,q,r,r)$.

\subsection{$h=0$ and the general formula}

When $\mathbf a=\mathbf 0$, Eq.~\eqref{eq:sm_pauli_fourth_moment_binary} simplifies to
\begin{align}
2^N\sum_{\mathbf x,\mathbf y,\mathbf z}\overline{|\Psi_{\mathbf x}|^2|\Psi_{\mathbf y}|^2|\Psi_{\mathbf z}|^2|\Psi_{\mathbf x\oplus\mathbf y\oplus\mathbf z}|^2}.
\label{eq:sm_zero_shift_start}
\end{align}
Extensive coincidences among the four bitstrings are now possible, and the counting differs from the $h\ge 1$ case. We count the triples $(\mathbf x,\mathbf y,\mathbf z)$ with prescribed charges
$|\mathbf x|=q_1$, $|\mathbf y|=q_2$, $|\mathbf z|=q_3$, and
$|\mathbf x\oplus\mathbf y\oplus\mathbf z|=q_4$. At each site, only the eight
even-parity local patterns $(0,0,0,0)$, $(0,0,1,1)$, $(0,1,0,1)$,
$(0,1,1,0)$, $(1,0,0,1)$, $(1,0,1,0)$, $(1,1,0,0)$, and $(1,1,1,1)$ can
occur. This immediately yields the generating polynomial
$1+uv+uw+uz+vw+vz+wz+uvwz$.
Hence
\begin{align}
N_{q_1q_2q_3q_4}:=[u^{q_1}v^{q_2}w^{q_3}z^{q_4}]\left(1+uv+uw+uz+vw+vz+wz+uvwz\right)^N
\label{eq:sm_N_def}
\end{align}
is precisely the desired counting coefficient.

The Haar average of Eq.~\eqref{eq:sm_zero_shift_start} splits into three
classes:

\begin{enumerate}
\item four distinct strings, counted by $N_{q_1q_2q_3q_4}$;
\item two-pair coincidences, weighted by $D_qD_r$;
\item complete fourfold coincidence, weighted by $D_q$.
\end{enumerate}

The coefficients $1$, $9$, $14$ are the multiplicities obtained by sorting the
$4!=24$ Wick contractions of Eq.~\eqref{eq:sm_sector_haar_moment} by their
equality pattern; they sum to $24$. We collect them as
\begin{align}
\mathcal C_N[p]&:=2^N\sum_{q_1,q_2,q_3,q_4=0}^{N}N_{q_1q_2q_3q_4}\Gamma_{q_1q_2q_3q_4}[p],
\label{eq:sm_C0_def}
\\
\mathcal D_N[p]&:=2^N\sum_{q,r=0}^{N}D_qD_r\,\Gamma_{qqrr}[p],
\label{eq:sm_D0_def}
\\
\mathcal E_N[p]&:=2^N\sum_{q=0}^{N}D_q\,\Gamma_{qqqq}[p].
\label{eq:sm_E0_def}
\end{align}
Then the zero-shift contribution is
\begin{align}
\mathcal F_N^{(0)}[p]=\mathcal C_N[p]+9\,\mathcal D_N[p]+14\,\mathcal E_N[p].
\label{eq:sm_F_h0_final}
\end{align}
Summing over all bit-flip sectors gives
\begin{align}
\mathcal F_N[p]:=\overline{\sum_{P\in\mathcal P_N}\left|\bra{\Psi_\infty}P\ket{\Psi_\infty}\right|^4}=\mathcal F_N^{(0)}[p]+\sum_{h=1}^{N}\binom{N}{h}\,\mathcal F_N^{(h)}[p].
\label{eq:sm_F_total}
\end{align}
Once each fixed-charge block is fully scrambled, the late-time stabilizer R\'enyi entropy depends on the initial state only through $p(q)$.

Using Eq.~\eqref{eq:sm_xi2_def}, the corresponding averaged stabilizer purity is
\begin{align}
\overline{\Xi_2^{(\infty)}}[p]=2^{-N}\mathcal F_N[p].
\label{eq:sm_xi2_ss_general}
\end{align}
Define the typical steady-state second stabilizer R\'enyi entropy as
\begin{align}
M_{2,\mathrm{typ}}^{(\infty)}[p]:=-\log_2 \overline{\Xi_2^{(\infty)}}[p]=N-\log_2 \mathcal F_N[p].
\label{eq:sm_M2_ss_general}
\end{align}
When the relevant sector dimensions are exponentially large the Haar distribution
concentrates sharply, so $M_{2,\mathrm{typ}}^{(\infty)}$ also gives the typical late-time value.

\subsection{Tilted product-state families}

\smallskip
\noindent\textit{Special point \texorpdfstring{$\theta=\pi/2$}{theta=pi/2}.}

At $\theta=\pi/2$, all three tilted product-state families considered in the
main text share the same sector weights,
\begin{align}
p_{\pi/2}(q)=\frac{D_q}{2^N}=\frac{1}{2^N}\binom{N}{q},
\label{eq:sm_p_pi_over_two}
\end{align}
since every computational-basis configuration occurs with equal modulus. The
global-Haar prediction therefore reduces to a one-parameter specialization of
Eq.~\eqref{eq:sm_M2_ss_general},
\begin{align}
M_{2,\mathrm{typ}}^{(\infty)}\!\left(N,\frac{\pi}{2}\right)=N-\log_2 \mathcal F_N[p_{\pi/2}],
\label{eq:sm_M2_pi_over_two_exact}
\end{align}

We do not have a comparably compact closed form for $\mathcal F_N[p_{\pi/2}]$
at finite $N$, but the large-$N$ asymptotics is straightforward. Substituting
$p(q)=D_q/2^N$ in Eq.~\eqref{eq:sm_F_total} gives
\begin{align}
\mathcal F_N[p_{\pi/2}]=4-\frac{12}{2^N}+O(4^{-N}),
\label{eq:sm_F_pi_over_two_asymptotic}
\end{align}
and therefore
\begin{align}
M_{2,\mathrm{typ}}^{(\infty)}\!\left(N,\frac{\pi}{2}\right)=N-2+\frac{3}{(\ln 2)\,2^N}+O(4^{-N}).
\label{eq:sm_M2_pi_over_two_asymptotic}
\end{align}
In particular, the steady-state stabilizer R\'enyi entropy is extensive with
unit slope and a universal offset of $-2$:
\begin{align}
\frac{M_{2,\mathrm{typ}}^{(\infty)}(N,\pi/2)}{N}=1-\frac{2}{N}+O\!\left(\frac{2^{-N}}{N}\right).
\label{eq:sm_M2_pi_over_two_density}
\end{align}
At $\theta=\pi/2$ the large-$N$ behavior is therefore simple: the magic
density approaches unity, and $M_2^{(\infty)}\to N-2$ with corrections
exponentially small in $N$.

\smallskip
\noindent\textit{Tilted ferromagnetic state.}

For the tilted ferromagnetic state
\begin{align}
\ket{\psi(\theta)}_{\rm TFS}
=
\bigotimes_{j=1}^{N}
\left(
\cos\frac{\theta}{2}\ket{0}
+
\sin\frac{\theta}{2}\ket{1}
\right),
\end{align}
it is convenient to define
 $c:=\cos(\theta/2)$ and $s:=\sin(\theta/2)$.
The charge-sector generating function is simply
\begin{align}
\sum_{q=0}^{N}p_{\rm TFS}(q)x^q=\left(c^2+s^2x\right)^N,
\end{align}
so the charge-sector probabilities are binomial:
\begin{align}
p_{\rm TFS}(q)=\binom{N}{q}\left(c^2\right)^{N-q}\left(s^2\right)^q.
\label{eq:sm_p_tfs}
\end{align}
Substituting Eq.~\eqref{eq:sm_p_tfs} into the general formula \eqref{eq:sm_M2_ss_general} gives the steady-state prediction
\begin{align}
M_{2,\mathrm{typ}}^{(\infty),{\rm TFS}}(N,\theta)=N-\log_2 \mathcal F_N[p_{\rm TFS}(\theta)].
\label{eq:sm_M2_tfs_ss}
\end{align}
All ingredients of $\mathcal F_N[p_{\rm TFS}]$ are now explicit: the sector
dimensions are binomials, the weights are given by Eq.~\eqref{eq:sm_p_tfs},
and the counting factors by Eqs.~\eqref{eq:sm_O_def} and \eqref{eq:sm_N_def}.

As a quick check, for $N=1$ we have $p_{\rm TFS}(0)=\cos^2(\theta/2)$ and $p_{\rm TFS}(1)=\sin^2(\theta/2)$. Evaluating Eq.~\eqref{eq:sm_M2_ss_general} gives
\begin{align}
\mathcal F_1[p_{\rm TFS}]=1+\cos^4\theta+\frac{3}{4}\sin^4\theta,
\label{eq:sm_tfs_one_qubit_F}
\end{align}
and therefore
\begin{align}
M_{2,\mathrm{typ}}^{(\infty),{\rm TFS}}(1,\theta)=1-\log_2\!\left(1+\cos^4\theta+\frac{3}{4}\sin^4\theta\right),
\label{eq:sm_tfs_one_qubit_M2}
\end{align}
matching direct one-qubit evaluation.

\smallskip
\noindent\textit{Tilted N\'eel and tilted domain-wall states.}

For even $N$, the tilted N\'eel state is
\begin{align}
\ket{\psi(\theta)}_{\rm TNS}
&=
\bigotimes_{j=1}^{N/2}
\left(
\cos\frac{\theta}{2}\ket{0}_{2j-1}
+
\sin\frac{\theta}{2}\ket{1}_{2j-1}
\right)
\nonumber\\
&\quad\otimes
\left(
-\sin\frac{\theta}{2}\ket{0}_{2j}
+
\cos\frac{\theta}{2}\ket{1}_{2j}
\right).
\end{align}
The charge generating function admits two equivalent forms.

The odd sublattice contributes $(c^2+s^2x)^{N/2}$, the even sublattice contributes $(s^2+c^2x)^{N/2}$, so
\begin{align}
\sum_{q=0}^{N}p_{\rm TNS}(q)x^q=\left(c^2+s^2x\right)^{N/2}\left(s^2+c^2x\right)^{N/2}.
\label{eq:sm_tns_genfun_factorized}
\end{align}
Alternatively, group the state into odd-even pairs. Each pair contributes charge $0$, $1$, or $2$ with weights
\begin{align}
c^2s^2,\qquad c^4+s^4,\qquad c^2s^2,
\end{align}
respectively, giving the equivalent pair form
\begin{align}
\sum_{q=0}^{N}p_{\rm TNS}(q)x^q=\left[c^2s^2+(c^4+s^4)x+c^2s^2x^2\right]^{N/2}.
\label{eq:sm_tns_genfun_pair}
\end{align}
Extracting the coefficient of $x^q$ gives
\begin{align}
p_{\rm TNS}(q)
&=
\sum_{k=\max(0,q-N/2)}^{\min(q,N/2)}
\binom{N/2}{k}\binom{N/2}{q-k}
\nonumber\\
&\quad\times
\left(c^2\right)^{N/2+q-2k}
\left(s^2\right)^{N/2-q+2k},
\label{eq:sm_p_tns}
\end{align}
matching the main text.

Substituting Eq.~\eqref{eq:sm_p_tns} into Eq.~\eqref{eq:sm_M2_ss_general} gives
\begin{align}
M_{2,\mathrm{typ}}^{(\infty),{\rm TNS}}(N,\theta)=N-\log_2 \mathcal F_N[p_{\rm TNS}(\theta)],
\label{eq:sm_M2_tns_ss}
\end{align}
again with every ingredient of $\mathcal F_N[p]$ explicit.

The same formula applies immediately to the tilted domain-wall state. As shown
in the main text,
\begin{align}
p_{\rm TDWS}(q)=p_{\rm TNS}(q)\qquad \text{for all } q,
\end{align}
and since Eq.~\eqref{eq:sm_M2_ss_general} depends on the initial state only
through $p(q)$, we obtain
\begin{align}
M_{2,\mathrm{typ}}^{(\infty),{\rm TDWS}}(N,\theta)=M_{2,\mathrm{typ}}^{(\infty),{\rm TNS}}(N,\theta).
\label{eq:sm_tdws_tns_same_ss}
\end{align}
In the global-$\mathrm{U(1)}$ Haar steady state, $M_2$ for the tilted N\'eel
and tilted domain-wall families therefore coincides; any difference at finite
depth reflects incomplete intra-sector scrambling.

\section{Initial Second Stabilizer R\'enyi Entropy of Tilted Product States}

For a single qubit, let
\begin{align}
\ket{\varphi(\theta)}
=
\cos\frac{\theta}{2}\ket{0}
+
\sin\frac{\theta}{2}\ket{1},
\label{eq:sm_local_tilted_spinor}
\end{align}
with Pauli expectations $\langle X\rangle=\sin\theta$, $\langle Y\rangle=0$, $\langle Z\rangle=\cos\theta$. The tilted ferromagnetic state $\ket{\psi(\theta)}_{\rm TFS}=\ket{\varphi(\theta)}^{\otimes N}$ is a product state, so $\Xi_2$ in Eq.~\eqref{eq:sm_xi2_def} factorizes into $N$ identical one-site contributions:
\begin{align}
\Xi_2\!\left(\ket{\psi(\theta)}_{\rm TFS}\right)
=
\left[
\tfrac{1}{2}\!\left(1+\sin^4\theta+\cos^4\theta\right)
\right]^N
=
\left[
1-\tfrac{1}{4}\sin^2(2\theta)
\right]^N,
\label{eq:sm_initial_xi2_tfs}
\end{align}
giving
\begin{align}
M_2\!\left(\ket{\psi(\theta)}_{\rm TFS}\right)
=
-N\log_2\!\left[
1-\tfrac{1}{4}\sin^2(2\theta)
\right].
\label{eq:sm_initial_m2_tilted}
\end{align}

The tilted N\'eel and tilted domain-wall states are also products of $\ket{\varphi(\theta)}$ and $\ket{\varphi'(\theta)}=-\sin(\theta/2)\ket{0}+\cos(\theta/2)\ket{1}$ in different sublattice orderings. The two spinors share identical $|\langle P\rangle|^4$ for all $P\in\{I,X,Y,Z\}$, so each site contributes the same factor and Eq.~\eqref{eq:sm_initial_m2_tilted} applies to all three families, reproducing Eq.~\eqref{eq:M2_main} in the main text.

\end{document}